\documentclass[twocolumn]{aa}
\usepackage{psfig}
\begin{document}


\title{The Type Ia Supernova 2001V in NGC 3987\footnote[1]{
Based on observations obtained at German-Spanish
Astronomical Centre, Calar Alto, operated by the Max-Planck-Institute 
for Astronomy, Heidelberg, jointly with the Spanish National 
Commission for Astronomy}}

\author{J. Vink\'o \inst{1,5} \and
I.B. B\'{\i}r\'o \inst{2} \and
B. Cs\'ak \inst{1,5} \and
Sz. Csizmadia \inst{3} \and
A. Derekas \inst{4,5} \and
G. F\H ur\'esz \inst{4,5} \and
Z. Heiner \inst{1,5} \and
K. S\'arneczky \inst{1,5,6} \and
B. Sip{\H o}cz \inst{5} \and
Gy. Szab\'o \inst{4,5,6} \and
R. Szab\'o \inst{3} \and
K. Szil\'adi \inst{1,5,6} \and
K. Szatm\'ary \inst{4}
}

\institute{Department of Optics \& Quantum Electronics, University of Szeged,
POB 406, Szeged, H-6701 Hungary \and
Baja Observatory, POB 766, Baja, H-6500 Hungary \and 
Konkoly Observatory of the Hungarian Academy of Sciences, POB 67, 
Budapest, H-1525 Hungary \and
Department of Experimental Physics, University of Szeged, D\'om t\'er 9., Szeged,
H-6720 Hungary \and
Guest Observer, Piszk\'estet\H o Station, Konkoly Observatory, Hungary \and
Visiting Astronomer, German-Spanish Astronomical Centre,
Calar Alto, Spain
}

\titlerunning{SN 2001V in NGC 3987}
\authorrunning{J. Vink\'o {\it et al.}}
\offprints{vinko@physx.u-szeged.hu}
\date{}

\abstract
{CCD photometry of the type Ia SN 2001V occured in the edge-on spiral galaxy NGC~3987
is presented. The observations made through Johnson-Cousins {\it BVRI} filters were collected
from Feb. 24 ($t = - 8$ days, with respect to B-maximum), 
up to May 5 ($t = +62$ days).
The light curves are analyzed with the revised Multi-Colour Light Curve Shape (MLCS)
method (\cite{riess2}) by fitting template vectors to the observed light curves
simultaneously. The reddening of SN~2001V is estimated to be $E(B-V)=0.05$ mag, while
the galactic component is $E(B-V) = 0.02$ mag (\cite{sfd}), suggesting that
part of the reddening may be due to the ISM  in the host galaxy. 
The $\Delta$ parameter in MLCS converged to $-0.47$ mag, indicating that this SN
was overluminous relative to the majority of Type Ia SNe. The inferred distance
to its host galaxy, NGC~3987, is $74.5 \pm 5$ Mpc, which is in good agreement with
recently determined kinematic distances, based on radial velocity corrected for
Virgo-infall and Hubble constant $H_0 = 65$ kms$^{-1}$Mpc$^{-1}$.}
  
\maketitle

\keywords{Stars: supernovae: individual: SN~2001V}

\section{Introduction}

SN~2001V was discovered spectroscopically by P.~Berlind at the F.~L.~Whipple
Observatory on Feb. 19, 2001 (\cite{iauc7585}). The supernova was immediately
classified as of Type Ia based on the \ion{Si}{ii} absorption trough 
at 6150 \AA. The blue continuum suggested that this SN was discovered 
before maximum light. This was also strengthened by the high expansion velocity
(14 000 kms$^{-1}$) derived from the center of the \ion{Si}{ii} line.
Because the light curves of Type Ia SNe are now frequently used to infer
distances to the host galaxies, such SNe that are identified before maximum are
very important, because i) the light curve can be sampled more effectively,
ii) the inferred distances are better constrained by the light variation around
maximum than at later phases. 

The host galaxy, NGC~3987 belongs to the small group centered on NGC~4005 (\cite{nilson},
\cite{greg}). This group is located within the medium-sized cluster Zw 127-10
(\cite{zw}). NGC~3987 is an edge-on, Sbc-type galaxy with remarkable central
dust lane. The IRAS point source 11547+2528 is located close to the optical center of
this galaxy, which was also detected in radio (\cite{condon}). 

SN~2001V is the first supernova discovered in this galaxy.
The SN occured in the outskirts, almost at the visible edge of
the host (Fig.1). The position of the SN as well as its blue colour at maximum
may suggest that the effect of interstellar extinction is probably
not very high for SN~2001V. Indeed, the galactic component of the reddening in
this direction is only $E(B-V)=0.02$ mag, according to \cite{sfd}.

There are two recent distance estimates for NGC~3987: \cite{mould} lists 
$\mu_0 = 33.82$ mag (58.1 Mpc) based on near-IR Tully-Fisher relation, 
while \cite{vandriel} gives  $v = 4361$ kms$^{-1}$ as a group-averaged radial
velocity corrected for Virgo-infall that results in $d = 67.1$ Mpc 
(adopting $H_0 = 65$ kms$^{-1}$Mpc$^{-1}$). Based on the average of
these distance estimates, the expected maximum brightness of a Type Ia SN 
is about 14.6 mag. According to the NASA/IPAC Extragalactic Database,
the redshift of NGC~3987 is $z = 0.015$. Thus, SN~2001V is a relatively
nearby, low-redshift SN. 

In the followings, details of the observations and data reductions are given,
then the results of the photometric analysis are presented and discussed. 

\begin{figure}
\leavevmode
\begin{center}
\psfig{file=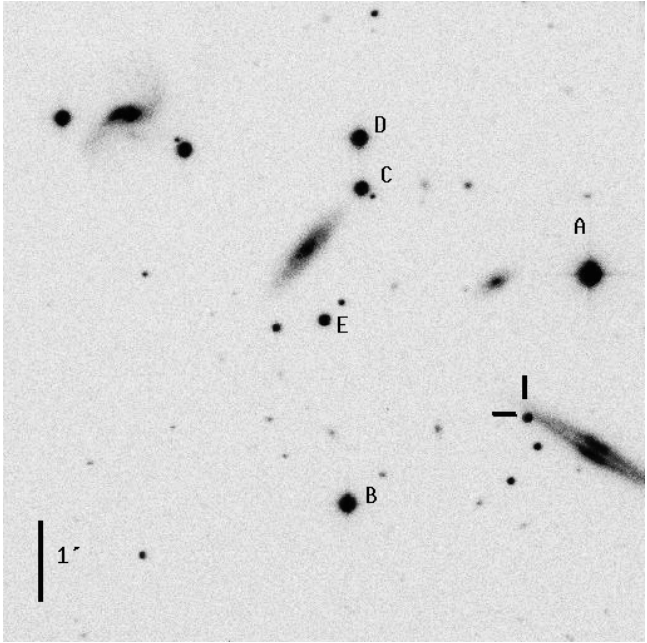,width=8.5cm}
\caption{The observed field of NGC~3987 taken with the Schmidt telescope
of Konkoly Observatory on Feb. 29, 2001. The position of SN~2001V is marked.
The comparison stars are labeled. North is up and east is to the left.}
\end{center}
\end{figure}
 
\begin{figure}
\leavevmode
\begin{center}
\psfig{file=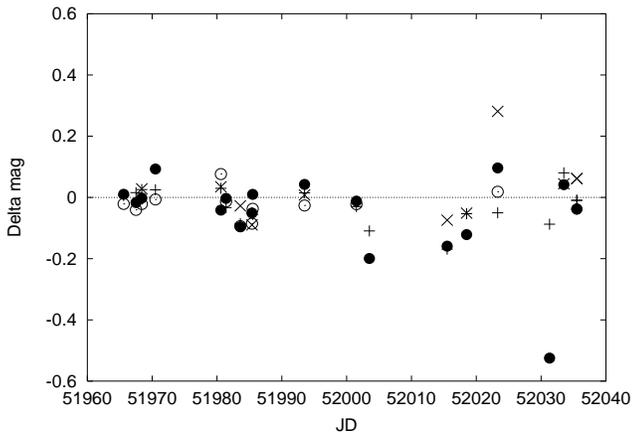,width=8.5cm}
\caption{The difference of magnitudes of SN~2001V obtained from
aperture- and PSF-photometry. Open circles: B-filter, filled circles:
V-filter, plus signs: R-filter, crosses: I-filter. }
\end{center}
\end{figure}

\section{Observations and data reduction}

The light variation of SN~2001V was followed on 18 nights, starting from
$t = -8$ days (with respect to $B-$maximum), 
extending up to $t = +62$ days.
The applied telescopes and detectors are listed in Table~1.
All data were collected through standard Johnson-Cousins $BVRI$ filters. 

\begin{table*}
\caption{Parameters of telescopes and detectors. The coloumns give
the name of the observatory, the telescope, the type of the CCD, 
the readout noise (in electrons), the number of pixels, 
the pixel scale (in arcsec/pixels), the total field of view (in arcmins), 
and the typical FWHM of the images (in pixels). 
An asterisk in the FWHM coloumn denotes 2x2 binning.}
\begin{tabular}{llccccccc}
\hline
\hline 
Code & Obs.& Tel. & CCD & RDN & Pixels & Scale & FoV & FWHM \\
\hline 
A&Konkoly Obs.&60 cm Schmidt&Photometrics& 14 $e^{-}$ & 1536x1024 & 1.00\" & 25.5x17\' & 3.3$^{~}$ \\
B&Konkoly Obs.&100 cm RCC&Photometrics& 6 $e^{-}$ & 1024x1024 & 0.29\" & 5x5\' & 3.1$^{*}$ \\
C&Konkoly Obs.&100 cm RCC&Wright EEV& 10 $e^{-}$ & 1200x800 & 0.36\" & 7x5\' & 3.2$^{*}$ \\
D&Szeged Obs.&40 cm Cass&SBIG ST-9& 13 $e^{-}$ & 512x512 & 0.70\" & 6x6\' & 3.0$^{*}$ \\
E&Baja Obs.&50 cm RCC&Apogee AP7& 18 $e^{-}$ & 512x512 & 1.29\" & 11x11\' & 3.3$^{~}$ \\
\hline
\end{tabular}
\end{table*}

The CCD-frames have been reduced in {\it IRAF}\footnote{{\it IRAF} is distributed by
NOAO which is operated by the Association of Universities for 
Research in Astronomy (AURA) Inc. under cooperative agreement with the National
Science Foundation}. First, the instrumental
magnitudes of the SN and the selected comparison stars (Fig.1) 
were derived with aperture photometry using the task {\it digiphot/apphot}.
The radius of the aperture was 6 pixels (about 2 x FWHM, see Table~1), while
the sky level was determined in a 5 pixel-wide annulus with inner radius of 
10 pixels. Because the SN is located close to the ``tip'' of NGC~3987, most
of the pixels in the annulus were not affected significantly by the light 
from the host galaxy. The sky level was determined by calculating the modal 
average (3 x median $-$ 2 x mean) of the intensities in the annulus. 
The calculations were done interactively, and the results were 
plotted on the screen and inspected visually in order to detect any obvious
systematic errors. The average sky level around the SN was always very close
to the background around the comparison stars, no clear systematic effect
could be found.

Second, the whole dataset have been re-reduced using PSF-photometry 
({\it digiphot/daophot}), as also advised by the referee of this paper.
PSF-photometry is a superior photometric method if the background 
strongly varies around the objects, and its removal is complicated.
Since SN~2001V is somewhat contaminated by its host, the
use of PSF-photometry may be useful to separate the light of the SN
from that of the galaxy.  This method requires bright
stars with high S/N to construct a good PSF. Because there are only a few
of such stars in the field of NGC~3987, and also the field of view of most of the
telescopes used in this study was quite small, a lot of frames contained
only 4 - 5 stars around the SN. Nevertheless, the PSF of each frame was determined
interactively. The analytic component of the PSF was approximated by
the built-in {\it penny2} function in {\it daophot}, 
but in most cases it was based
on only 2-3 stars. The frames made with the Schmidt-telescope (Table~1) contained
much more field objects, but the PSF on these frames showed significant 
positional dependence. Therefore, a second-order variable PSF model has
been constructed for these pictures, while constant PSF was determined for
all other frames. Individual sky levels were calculated for all objects,
and the background was subtracted iteratively during the fitting of the
PSF in {\it daophot/allstar}. Then, the residuals were examined visually
on the subtracted frames. The stars as well as the SN were adequately 
removed from most frames,
but slight residuals were present at the position of the brightest stars
on some frames. This was probably caused by the uncertainty of the PSF due
to the small number of PSF-stars. 

Differential magnitudes of SN~2001V have been computed using the comparison
stars labeled in Fig.1 (see also Table~2 below). After transforming
them to the standard system (\cite{2ke}), the brightness of the SN was 
calculated from the calibrated {\it BVRI} magnitudes of each comparison 
star (see below). Then, the SN magnitudes belonging to the same frame were
averaged.  

In order to search for any systematic effect caused by the reduction procedure,
we have compared the SN magnitudes from the aperture- and PSF-photometry.
Fig.2 presents the aperture minus PSF-magnitudes in all filters 
as a function of time. It is apparent that most of the differences are
within $\pm 0.1$ mag. 
Naturally, the differences are higher at later phases, when the
SN became fainter, but there is no visible systematic trend in the
data. We conclude that both the aperture- and PSF-photometry of
SN~2001V presented here is affected by approximately the same
amount of random errors at the $\pm 0.1$ mag level, mainly due
to the technical limitations of the applied instruments (lower S/N,
small field of view, very few PSF-stars). In order to reduce the
uncertainties introduced by the reduction method, the SN magnitudes
resulted from both aperture- and PSF-photometry were averaged, and
these magnitudes were accepted as the final result.

At first, the magnitudes of the comparison stars were calibrated via 
Landolt standards,
observed at Calar Alto Observatory with the 1.2 m Cassegrain, under photometric
conditions on Aug.11, 2001. The reliability of this dataset
have been checked by using the calibrated photometry of some of the field
stars made by the CfA Supernova Group at the F.~L.~Whipple Observatory.
Because the comparison stars used in this study, and those adopted by
the CfA-group were different, we have re-computed the magnitudes of
our comparison stars using the CfA-dataset as secondary standards on
a set of $BVRI$ frames collected with the Schmidt telescope of Konkoly
Observatory on March 14, 2001. After performing aperture photometry
(with the same parameters described above) and standard transformation,
the ``Calar Alto'' calibration and the ``CfA/Konkoly'' calibration 
was found to be fully consistent. The differences
between the two datasets were within $\pm 0.02 ~-~ 0.03$ mag.  
Finally, the results
from the CfA/Konkoly calibration was adopted, in order to keep maximal 
consistency with data of SN~2001V from other observatories. The final 
standard magnitudes of the comparison stars are collected in Table~2. The 
uncertainties reflect the statistical errors given by {\it daophot}
and also the magnitude differences between the two calibrating datasets.

The calibrated standard magnitudes of SN~2001V are listed in Table~3.
The estimated errors of each point are given in parentheses. These
uncertainties were calculated as 
$\sigma^2 = \sigma_p^2 + \sigma_m^2 + \sigma_t^2$,
where $\sigma_p$ is the statistical error given by {\it daophot},
$\sigma_m$ is the error due to the photometric method (assumed to be
equal to the difference between the aperture- and PSF-photometry, plotted
in Fig.2) and $\sigma_t$ is the error introduced by the standard transformation
(adopted as 0.03 mag for each data).

\begin{table}
\caption{Comparison stars in the field of NGC~3987}
\begin{tabular}{lcccc}
\hline
\hline
Star & $V$ & $B-V$ & $V-R$ & $V-I$ \\
\hline
A & 10.67 (0.01) & 0.59 (0.02) & 0.44 (0.02) & 0.71 (0.02)\\
B & 12.23 (0.02) & 0.84 (0.02) & 0.48 (0.02) & 0.95 (0.02)\\
C & 13.00 (0.02) & 0.59 (0.03) & 0.36 (0.02) & 0.69 (0.03)\\
D & 12.25 (0.02) & 0.43 (0.05) & 0.28 (0.02) & 0.57 (0.02)\\
E & 14.77 (0.03) & 0.85 (0.06) & 0.49 (0.04) & 0.89 (0.05)\\
\hline
\end{tabular}
\end{table}

\begin{table*}
\caption{Photometry of SN~2001V. The uncertainties are given in parentheses.
See Table~1 for telescope codes.}
\begin{tabular}{lccccc}
\hline
\hline
JD & $B$ & $V$ & $R$ & $I$ & Tel. \\
\hline
51965.6 & 15.04(0.037) & 15.04(0.033) & 15.04(0.033) & -- & A \\
51967.5 & 14.82(0.051) & 14.84(0.035) & 14.83(0.036) & -- & A \\
51968.4 & 14.72(0.037) & 14.73(0.030) & 14.75(0.041) & 14.91(0.042) & A \\
51970.5 & 14.79(0.035) & 14.72(0.098) & 14.74(0.043) &  -- & E \\
51980.6 & 15.10(0.093) & 14.89(0.061) & 14.72(0.046) & 15.15(0.052) & D \\
51981.4 & 14.92(0.040) & 14.72(0.030) & 14.68(0.045) &  -- & E \\
51983.6 & 15.22(0.098) & 14.86(0.102) & 14.89(0.093) & 15.25(0.060) & D \\
51985.4 & 15.40(0.092) & 15.00(0.060) & 14.94(0.058) & 15.28(0.095) & D \\
51985.5 & 15.20(0.050) & 14.89(0.032) & 14.86(0.064) &  -- & C \\
51993.5 & 15.93(0.040) & 15.26(0.052) & 15.23(0.034) & 15.31(0.031) & A \\
51997.5 & 16.51(0.050) & 15.52(0.040) & 15.26(0.050) & 15.27(0.055) & C \\
52001.5 & 16.89(0.043) & 15.79(0.032) & 15.37(0.034) &  -- & E \\
52003.5 &  -- & 15.78(0.202) & 15.40(0.115) & -- & E \\
52015.5 &  -- & 16.49(0.163) & 15.99(0.172) & 15.79(0.080) & D \\
52018.5 &  -- & 16.52(0.132) & 16.20(0.072) & 15.79(0.090) & D \\
52023.3 & 17.53(0.036) & 16.55(0.101) & 16.45(0.060) & 16.15(0.100) & A \\
52031.3 &  -- & 17.04(0.120) & 16.66(0.095) & 16.42(0.120) & D \\
52033.5 &  -- & 17.01(0.110) & 16.81(0.132) & 16.79(0.110) & B \\
52035.5 &  -- & 17.06(0.111) & 16.76(0.104) & 16.70(0.121) & B \\
\hline
\end{tabular}
\end{table*}

The observations listed in Table~3 were compared with the $R-$band CCD photometric 
data of K.Hornoch (\cite{hor}) and found reasonable agreement 
within $\pm 0.1$ mag.
More recently \cite{mandel} published $B_{\rm max} = 14.64 \pm 0.03$ mag 
based on a more extensive photometric dataset. 
This agrees very well with our template fitting 
(see next section) resulting in $B_{\rm max} = 14.60 \pm 0.16$ 
(the larger uncertainty
is due to the lack of data around maximum in our photometry). 
This agreement
may give some support to our belief that the data in Table~3 probably 
do not suffer from large systematic errors.

\section{Results}

\begin{figure}
\leavevmode
\begin{center}
\psfig{file=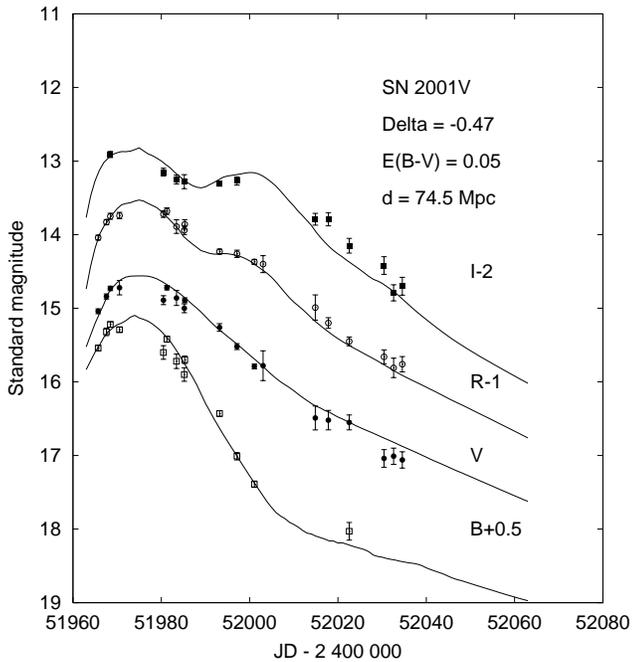,width=8.5cm}
\caption{Light curves of SN~2001V in {\it B, V, R, I} filters 
(the {\it B, R, I} data were shifted for better visibility). 
The lines are the template light curves fitted by the MLCS method
(see text).}
\end{center}
\end{figure}

The {\it BVRI} light curves were analyzed by the Multi-Colour Light
Curve Shape (MLCS) method of \cite{riess2} (the SN template
vectors were kindly provided by A. Riess). The timescale of the
observations was divided by $1+z$ = 1.015 to remove the effect
of time dilation. The value of $K$-correction in the $V$ band at
this redshift varies between $-0.01$ and $0.03$ mag (\cite{hamuy1}),
which is definitely smaller than the uncertainty of our measurements,
thus, it was not taken into account. 
The maximum absolute magnitude of the fiducial light curve was 
adopted as $M^{\rm max}_V = -19.46$ mag (\cite{rich}).
The fitting was computed
simultaneously to all light curves. Details of the procedure
are described in \cite{2ke}. The minimum of the $\chi^2$ was found
at the following parameters: $T_0(B)$ = JD 2451973.0 $\pm$ 0.2 
(fiducial $B$-maximum),
$E(B-V)$ = $0.05 \pm 0.02$ mag, $\mu_0$ = $34.36 \pm 0.14$ mag and 
$\Delta$ = $-0.47 \pm 0.04$ mag. The given uncertainties are formal
errors corresponding to the geometry of the $\chi^2$ function around
minimum. The fitted light curves are shown in Fig.3. Note, that the
$T_0(B)$ parameter given here corresponds to the maximum of the 
{\it fiducial}
light curve (with $\Delta = 0.0$), the actual maximum of SN~2001V 
in $B$ occured one day later, at JD 2451974. 

\begin{figure}
\leavevmode
\begin{center}
\psfig{file=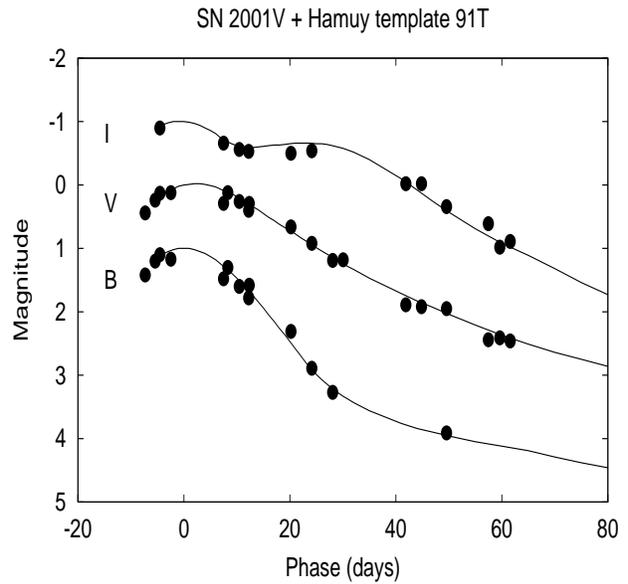,width=8.5cm,height=8cm}
\caption{Comparison of light curves of SN~2001V (symbols) and 
SN~1991T (continuous lines).}
\end{center}
\end{figure}

The parameters above imply some interesting properties of SN 2001V. 
Comparing the $E(B-V)$ colour excess with the reddening due to Milky Way
dust ($E(B-V)_{\rm gal} = 0.02$, \cite{sfd}), it is visible that the
ISM in the host galaxy contributes to the dust absorption.
Since NGC~3987 is an edge-on galaxy with strong central dust lane, it
is not an unexpected result. The host-galaxy reddening could have been even
more severe if the SN had occured closer to the central plane of NGC~3987,
like e.g. in the case of SN~1986G (\cite{phil1}), or, more recently
SN~2002cv in NGC~3190 (\cite{iauc7911}). 
   
The reddening was also estimated using the ``Lira-Phillips law'' (\cite{phil2})
as an independent check of the value given by the MLCS method. This 
method uses the apparent homogeneity of the unreddened $(B-V)_0$ colours of
SNe Ia in the $30  < t_V < 90$ phase interval ($t_V$ is the phase 
measured from $V-$maximum). Unfortunately, there is only one observed $(B-V)$
data point in Table~3 that can be used for this purpose ($(B-V) = 0.98 \pm 0.10$
mag). 
The moment of $V-$maximum was estimated from the template light curves fitted by the
MLCS method. It occured at $T_0(V)$ = 2451976 $\pm 1$ JD. 
After correcting for time dilation, the phase of the observed $(B-V)$ point 
was found to be $t_V = 46.3 ~ \pm 1$ days. The relation
$$ (B-V)_0 ~=~ 0.725 ~-~ 0.0118 ( t_V - 60) $$
given by \cite{phil2} resulted in $(B-V)_0 = 0.89 \pm 0.02$. This implies
$E(B-V) = 0.09$ mag, which is not too far from the
value given by MLCS. However, its uncertainty is probably higher, 
because this estimate is based on a single observation,
and the ``Lira-Phillips law'' itself may have an intrinsic uncertainty at the
$0.1$ mag level (\cite{li1}). On the other hand, the similarity of
the reddening values given by two independent methods may mean that
the estimated $E(B-V)$ is close to the real value. 

Adopting $E(B-V) = 0.05$ mag, the total absorption in
the $B$, $V$, $R$ and $I$ bands is $A_B = 0.21$, $A_V = 0.16$, $A_R = 0.13$,
$A_I = 0.09$ mag, respectively, using the galactic reddening
law given by \cite{sfd}.  

The luminosity parameter $\Delta = -0.47$ implies that SN~2001V is a super-luminous
Type Ia SN, relative to the majority of such SNe. 
Super-luminous SNe are often referred as SN~1991T-, or
SN~1999aa-like events (see \cite{li2}).
The distinction between different subtypes of SNe Ia is usually based on
spectroscopic properties, e.g. the strength of the \ion{Si}{ii} trough
at 6150 \AA. Overluminous SNe Ia have weaker \ion{Si}{ii} 6150 \AA~line, 
and the \ion{Fe}{iii} lines are strong. In addition, SN~1999aa-like
events usually show strong \ion{Ca}{ii} H \& K lines, which are weak in
SN~1991T-like SNe. These spectroscopic characteristics correlate with
photometric properties: the light curves of SN~1991T-like SNe declines
more slowly, and they are bluer in maximum than normal SNe Ia. Thus,
overluminous SNe may be recognizable from photometry, e.g. via the
MLCS-method. 

We do not have spectroscopic data of SN~2001V at our disposal, but
two spectra of SN~2001V, obtained around maximum, are available in graphical
form from the CfA-website 
\footnote{http://cfa-www.harvard.edu/cfa/oir/Research/supernova}. As the
referee of this paper, Dr. Weidong Li, pointed out, these spectra are
very similar to that of SN~1999aa: broad, but less prominent \ion{Si}{ii}
6150 \AA~line, strong \ion{Fe}{iii} and \ion{Ca}{ii} H \& K lines. 
These spectral features fully support our conclusion (based on photometry), 
that SN~2001V is an overluminous SN Ia.

Note, however, that it is not clear whether the higher luminosity is necessarily
connected with the spectroscopic peculiarities. There are examples, such as
SN~1992bc (\cite{riess2}), that were super-luminous, but otherwise showed normal 
spectra around maximum. Thus, the overluminosity of some Type Ia SNe might be 
due to e.g. statistical fluctuation of the ejected $^{56}$Ni mass that powers the
light curve, rather than a difference in the explosion mechanism or the physical
state of the progenitor. Recently \cite{rich} found that the
maximum brightness $M_B$ of SNe Ia has a Gaussian distribution with 
$\sigma = 0.56$ mag (corrected for the effect of extinction).   

From yet unpublished photometry, recently \cite{mandel} concluded that
SN~2001V was a ``normal'' Type Ia event, with light
and colour curves very similar to SN~1990N. On the other hand, 
they also determined the initial decline rate, and found
$\Delta m_{15}(B) = 0.99 \pm 0.05$. This decline rate may suggest an
overluminous, rather than a ``normal'' SN Ia (e.g. \cite{hamuy2}). 
Although our photometry has less accuracy and phase coverage, 
we also attempted to estimate this parameter and found 
$\Delta m_{15}(B) = 0.9 \pm 0.1$.
According to the correlation between $\Delta m_{15}(B)$ and $\Delta$
(e.g. \cite{riess2}), this is consistent with the $\Delta$ parameter
given by the MLCS-method. 
Furthermore, Fig.4 shows the essential similarity of the light 
curves of SN~2001V with that of SN~1991T 
($\Delta m_{15}(B) = 0.94$, \cite{hamuy2}). 

Thus, all available data 
consistently support the conclusion that SN~2001V was an intrinsically bright
SN Ia. At maximum, it was brighter than the fiducial Type Ia SN by about 
0.5 mag. Adopting the parameters given by the MLCS-method, 
the inferred absolute magnitudes of SN~2001V at maximum are

\begin{center}
$M^{\rm max}_B$ = $-19.97 ~\pm~ 0.15$ mag \\
$M^{\rm max}_V$ = $-19.96 ~\pm~ 0.15$ mag \\
$M^{\rm max}_R$ = $-19.96 ~\pm~ 0.15$ mag \\
$M^{\rm max}_I$ = $-19.63 ~\pm~ 0.15$ mag \\
\end{center}

The true distance modulus, $\mu_0 = 34.36 \pm 0.14$ mag, corresponds to
$74.5 \pm 5$ Mpc. The uncertainty given here is the formal error of
the fitting, and does not include possible systematic effects, 
such as the zero-point of the SN distance scale. 
The MLCS distance marginally agrees with the kinematic distance of NGC~3987
corrected for Virgo-infall (67 Mpc, see Sect.1), the difference is about
$1.5 ~ \sigma$. 
The Tully-Fisher distance (58 Mpc, Sect.1) is about 20 percent shorter.
The TF distance modulus ($\mu_0(TF) = 33.82$, \cite{mould})
can be brought into agreement with the MLCS distance by adding 0.54
mag to the previous one, in accord with the suggestion by Shanks
(\cite{shanks}) for matching the TF and SN distance scales.

\section{Summary}

The results of this paper can be summarized as follows.

\begin{enumerate}

\item{We presented {\it BVRI} photometry of SN~2001V starting from
8 days before maximum light and extending up to 60 days past maximum.}

\item{The reddening was estimated by applying the MLCS template fitting 
method for all light curves. This resulted in 
$E(B-V) = 0.05 \pm 0.02$ mag, indicating
that there was significant extinction in the host galaxy affecting the
light from SN~2001V. The empirical ``Lira-Phillips law'' gave similar,
but slightly higher colour excess ($E(B-V) = 0.09$ mag). Its
uncertainty is probably higher, because it is based only a single observation.
Therefore, the result from the MLCS-method was accepted as final. 
}

\item{The light curve analysis indicates that SN~2001V was an over-luminous
SN by about 0.5 mag relative to the fiducial SN Type Ia. This is consistently
supported by the overall similarity of the light curves and the $\Delta m_{15}(B)$
decline rate with those of SN~1991T, and also the publicly available spectra
of SN~2001V. The spectral features suggest that SN~2001V may be a SN~1999aa-like
object (\cite{li2}).}

\item{The SN distance was inferred from the MLCS method and found to be 
74.5 $\pm 5$ Mpc. This marginally agrees with the previous kinematic distance, 
but about 20 percent longer than the Tully-Fisher distance of NGC~3987.}

\end{enumerate}

\begin{acknowledgement}
This work was supported by Hungarian OTKA Grants No. T032258, T034615, 
the ``Bolyai J\'anos'' Research Scholarship to JV, the
OM FKFP 0010/2001 grant from Hungarian Ministry of Education, 
Pro Renovanda Cultura Hungariae Foundation and MTA-CSIC Joint Project No.15/1998
from Hungarian and Spanish Academy of Sciences.
We acknowledge the permission by Prof. A. Riess to use the MLCS template 
vectors. We thank the referee, Dr. Weidong Li, for a lot of critical 
comments and suggestions that helped us to improve the paper. We are grateful
to Dr. T. Matheson and the Supernova Group led by Prof. R. Kirshner
at the Harvard-Smithsonian Center for Astrophysics for providing
their calibration data prior to publication.
Thanks are also due to Dr. L.L.Kiss for his kind help during
some observing sessions carried out at Szeged Observatory.
The NASA Astrophysics Data System, the SIMBAD and NED databases and the
Canadian Astronomy Data Centre were frequently used to access data and
references. The availability of these services are gratefully acknowledged.
\end{acknowledgement}

\end{document}